\documentclass[prb,twocolumn,showpacs]{revtex4}
\usepackage{graphicx}

\begin{document}

\title{Prediction of Ferromagnetic Correlations in Coupled Double-Level Quantum Dots}

\author{G. B. Martins}
%\email[Author to whom correspondence should be addressed at: ]{martins@magnet.fsu.edu}
\affiliation{Department of Physics, Oakland University, Rochester, MI 48309}
\author{C. A. B\"usser}
\author{K. A. Al-Hassanieh}
\author{A. Moreo}
\author{E. Dagotto}
\affiliation{Department of Physics, University of Tennessee, Knoxville, TN 37996}

\begin{abstract}
Numerical results are presented for transport properties of two coupled double-level 
quantum dots. The results strongly suggest that under appropriate circumstances the dots 
develop a novel ferromagnetic correlation at quarter-filling (one electron per dot). In the strong 
coupling regime (Coulomb repulsion larger than electron hopping) and with the inter-dot tunneling 
larger than the tunneling to the leads, an S=1 Kondo resonance develops in the density of states, 
leading to a peak in the conductance. A qualitative ``phase diagram'', incorporating the new FM 
phase, is presented. It is 
also shown that the conditions necessary for the ferromagnetic regime are less restrictive than 
naively believed, leading to its possible experimental observation in real quantum dots.
\end{abstract}

\maketitle

Strongly correlated electronic systems, such 
as high-T$_{\rm c}$ cuprates, heavy fermions, and manganites, 
display a variety of nontrivial collective states, 
which are difficult to analyze 
due to the many-body character of the interactions, 
and the difficulties in experimentally {\it controlling} 
the parameters determining these interactions. These 
problems are severe in 
materials that spontaneously grow in particular 
structures and patterns, with several effects (lattice, 
spin, charge, orbital) in direct competition. 
Therefore, the observation of a 
celebrated many-body effect, the Kondo effect, in a 
single quantum dot (QD) \cite{Goldhaber1} has captured the attention 
of the strongly correlated community. It is conceivable 
that the most interesting states that are spontaneously 
stabilized in some materials - and are very difficult to control - 
could instead be artificially created 
in a man-made structure. In this framework, a natural first-step 
is to analyze coupled QDs. In fact, the two-impurity Kondo 
problem - extensively 
studied since the 80's\cite{jayap} - can now be realized 
in a real system. Moreover, 
recent investigations have reported  antiferromagnetic (AF) correlations 
between two single-level coupled QDs, in competition 
with Kondo correlations \cite{twoqd,twoqdexp,carlos1}. As a consequence, it is 
now clear that two of the most remarkable magnetic 
states known to exist in spontaneously grown materials - 
the Kondo and AF states - have already found realizations in 
the context of QDs.
However, the other dominant magnetic state of some materials 
- the ferromagnetic (FM) state - 
has comparatively received much less attention \cite{craig}. 
For the dream of artificially replicating collective states 
using QDs to be fulfilled, a realization of FM states 
must be achieved. The lack of attention to FM states in QDs 
should not be surprising in view of the 
physics of FM materials, such as manganites. Here, the FM 
state is reached by removing electrons (doping)  
from an AF state. Under the 
constraint of having one particle per level (1/2-filling), and only one 
level active per QD as in most previous investigations, 
the double-exchange\cite{zenner} generated FM 
state {\it cannot} be realized. To reach a FM state, 
more levels need to be active, resulting in 
{\it less} than one electron per level. 

In this Letter, clear 
evidence is presented for the development of {\it ferromagnetic} 
correlations between two {\it double-level} QDs \cite{multi1}: 
at 1/4-filling (one electron per dot), two coupled
double-level QDs form a triplet state. Coupling this state
to ideal metallic leads produces a Kondo resonance and a peak
in the conductance.  The results do not appear to be restricted
to only a pair of QDs, but they seem valid for larger QD arrays.
Basically here {\it it is reported a realization of the double-exchange
mechanism using QDs}. Although the above mentioned 
effect is stronger if the appropriate intra-dot inter-level many-body 
interactions are added to the Hamiltonian, 
it is important to stress that these interactions are 
not necessary: considering just an intra-level Coulomb repulsion (Hubbard $U$) is enough to 
obtain qualitatively the same results, opening the possibility for the FM regime to be 
experimentally observable.
\begin{figure}[h]
\centering
\includegraphics[width=5.0cm]{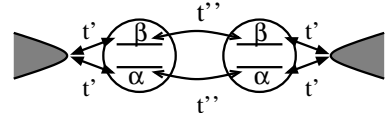}
\caption{ Schematic representation of the two double-level QDs system and the hopping 
amplitudes of Eqs. (1) and (3).
}
\label{fig1}
\end{figure}
Figure 1 schematically depicts the system analyzed here and introduces the labelling for the different 
tunneling parameters $t^{\prime}$ and $t^{\prime \prime}$. 
To model this system, the impurity Anderson Hamiltonian 
that describes the two QDs with two levels (denoted $\alpha$ and $\beta$) is given by
%\begin{flushright}
\begin{eqnarray}
H_{\rm d}=\sum_{i=1,2} \left\{ \sum_{\sigma;\lambda=\alpha,\beta}  
U n_{i\lambda \sigma} n_{i\lambda \bar{\sigma}} + 
\sum_{\sigma \sigma'} \left[U' n_{i\alpha \sigma} n_{i\beta \sigma'} - \right. \right.  \nonumber \\ 
\left. J c^{\dagger}_{i\alpha \sigma} c_{i\alpha \sigma'} c^{\dagger}_{i\beta \sigma'} 
c_{i\beta \sigma} \right] + \sum_{\sigma} \left[ V_g n_{i\alpha \sigma} + \right. \nonumber \\ 
\left. \left. (V_g + \Delta V) n_{i\beta \sigma} \right] \right\} +  \sum_{\sigma,\lambda=\alpha \beta} 
    t'' \left[ c^{\dagger}_{1\lambda \sigma} c_{2\lambda \sigma} + \mbox{h.c.} \right] ,
\end{eqnarray}
%\end{flushright}
where the first term represents the usual Coulomb repulsion between
electrons in the same level (considered equal 
for both levels). The
second term represents the Coulomb repulsion between electrons in
different levels (the $U'$ notation is borrowed from standard many-orbital
studies in atomic physics). The third term represents the Hund coupling 
($J>0$) that favors the alignment of spins and 
the fourth term is the energy of the states 
regulated by the gate voltage $V_g$. 
To decrease the number of free parameters, all the calculations presented here 
assume the following relations: $U^{\prime}=2U/3$ and $J=U-U^{\prime}$.
As discussed later, the main result in this Letter does not depend on the
specific values taken by $U^{\prime}$ and $J$.
Note that $\alpha$ and $\beta$ are separated by $\Delta V$,
and by modifying this parameter an interpolation between one- and two-level
physics can be obtained. The last term represents the inter-dot coupling, with 
matrix element $t^{\prime \prime}$. For simplicity, we assume 
that there is no hopping between levels $\alpha$ and $\beta$. 
The dots are connected to the leads (represented by semi-infinite chains)
by a hopping term with amplitude $t'$, while $t=1$ 
is the hopping amplitude in the leads (and energy scale). More specifically,
\begin{eqnarray}
H_{\rm leads} &=& t \sum_{i \sigma} \left[ c_{l i\sigma}^{\dagger} c_{l i+1\sigma}
 +  c_{r i\sigma}^{\dagger} c_{r i+1\sigma} +\mbox{h.c.} \right], \\
H_{\rm int} &=& t' \sum_{\sigma, \lambda=\alpha,\beta} \left[ 
 c_{1 \lambda \sigma}^{\dagger} c_{l 0\sigma} + c_{2 \lambda \sigma}^{\dagger} c_{r 0\sigma} 
 + \mbox{h.c.} \right] ,
\end{eqnarray}
where $c_{l i\sigma}^{\dagger}$ ($c_{r i\sigma}^{\dagger}$ ) creates electrons at
site $i$ with spin $\sigma$ in the left (right) contact. Site ``0'' is
the first site at the left of the left dot and at the right of the right dot, for each half-chain.
The total Hamiltonian is $H_{\rm T} = H_{\rm d} + H_{\rm leads} + H_{\rm int}$.
Note that for $V_g$=$-U/2 -U' +J/2 - \Delta V/2$, the Hamiltonian
is particle-hole symmetric.
Using the Keldysh formalism \cite{Meir-cnd}, the conductance through this system can be written as\cite{note1}
$\sigma =  \frac{e^2}{h} |t^2G_{lr}(E_{\rm F})|^2  \left[ \rho(E_{\rm F}) \right]^2$. 
In practice, a cluster containing the interacting dots and a few sites of the leads is solved 
exactly, the Green functions are calculated, and the leads are incorporated through a Dyson Equation 
embedding procedure (details of the 
embedding have been provided elsewhere\cite{carlos1,interfere}). In Figs. 
2-4, the cluster used involved the two QDs plus one lead site at left and right.
\begin{figure}[h]
\centering
\includegraphics[width=8cm]{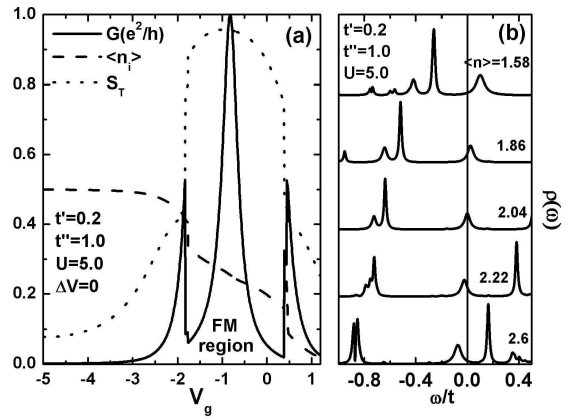}
\vspace{-0.5cm}
\caption{ Results for the strong inter-dot tunneling regime ($t^{\prime \prime}/t^{\prime} \gg 1$)
showing the presence of a {\it ferromagnetic} state 
at 1/4-filling and concomitant Kondo effect: (a) Conductance (solid line), charge occupancy 
per level and per spin (dashed line),
and total spin $S_{\rm T}$ in the four levels of the dots (dotted line) vs. gate potential $V_g$ for
two coupled double-level QDs. The values for the parameters
are indicated. Note that the maximum in the conductance corresponds to a spin 1 Kondo peak that occurs when there is
exactly one electron per dot and the total spin $S_{\rm T}\approx 1.0$.
The other two peaks occur when there are 1 and 3 electrons in the two-dots system,
corresponding to $S_{\rm T}=1/2$ Kondo resonances. 
(b) DOS indicating the presence of a Kondo resonance
associated with the $S_{\rm T}=1$ Kondo peak in (a). The numbers on
the right side indicate the total number of electrons inside
the two quantum dots. $V_g$ takes the values $0.0$, $-0.5$, $-0.83$, $-1.12$, $-1.5$ from the top
to the bottom curve.
The Fermi level is located at $\omega/t=0.0$.
}
\label{fig2}
\end{figure}

In Fig. 2a, results for conductance (solid line) at strong inter-dot tunneling ($t^{\prime \prime}/t^{\prime} 
\gg 1$) are shown. The main feature 
displayed is the peak at $V_g \approx -1.0$ (at and near 1/4-filling). The occupancy for this value of $V_g$ is 
approximately 1 electron per dot (dashed line $\times 4$) and the total spin $S_{\rm T}$ of the four levels 
\cite{note2}(dotted line) is $\approx 1.0$. The smooth charging of the levels as the gate potential 
decreases (in the peak region) indicates a possible Kondo regime. This is confirmed in Fig. 2b, 
where the density of states (DOS) close to the Fermi level is displayed as the gate potential varies 
from $0.0$ to $-1.5$ (top to bottom). Through this variation of $V_g$ the two dots are charged with one additional 
electron (the total mean charge varies from $\approx 1.6$ to $2.6$). 
One can clearly see a Kondo resonance pinned to the Fermi level.
For lower values of the gate potential (in the region at and near 1/2-filling, with 
2 electrons per dot) the conductance is drastically reduced and the total spin $S_{\rm T}$ inside 
the dots reaches its minimum value, indicating the formation of a global singlet state. Calculations 
of the total spin in each dot indicate that this singlet state is formed by the AF 
coupling of two spins $S=1$. A description of how this picture changes as the 
$t^{\prime \prime}$ decreases is shown in Fig. 3a, 
where results are shown for five different values of $t^{\prime \prime}$. 
\begin{figure}[h]
\centering
\includegraphics[width=8cm]{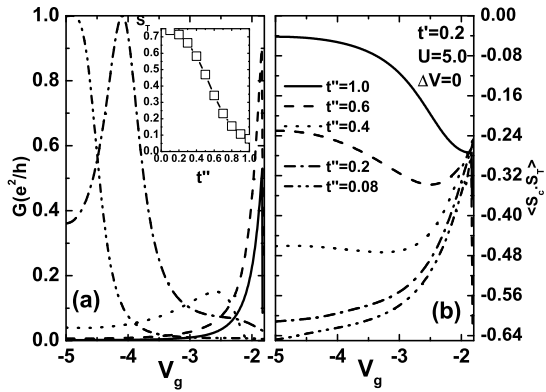}
\vspace{-0.5cm}
\caption{(a) Conductance $G$ vs. gate potential $V_g$ for different
values of inter-dot tunneling ($t^{\prime \prime}=1.0, 0.6, 0.4, 0.2,$ and $0.08$, 
see convention in (b)).
$V_g$ varies across the region where the fourth electron is charged into the
double-dot system. Note that the conductance at $V_g=-5.0$ gradually
increases from zero (at strong tunneling, $t^{\prime \prime}=1.0$) to the 
maximum value (at weak tunneling, $t^{\prime \prime}=0.08$). This variation
indicates a transition from two $S=1$ spins (in each dot) forming a global singlet
to two uncorrelated $S=1$ spins, each forming a Kondo resonance with the lead
conduction electrons. The inset shows the variation with $t^{\prime \prime}$ of the total spin
inside the dots. (b) Variation
with $t^{\prime \prime}$ of the spin-spin correlation between the total
spin in the double-dot and a conduction electron located in the first site
of the leads.
}
\label{fig3}
\end{figure}
The conductance at the particle-hole symmetric point, $V_g=-5.0$, 
varies from zero, for $t^{\prime \prime}=1.0$, to 1.0 (in units of $e^2/h$), 
for $t^{\prime \prime}=0.08$.
Fig. 3b shows how the Kondo correlation (between the total spin in the dots and a conduction 
electron in the first site of one of the leads) evolves from a negligible value for 
$t^{\prime \prime}/t^{\prime} \gg 1$ to a large value ($\approx -0.65$) 
for $t^{\prime \prime}/t^{\prime} < 1$. The inset of Fig. 3a displays the change of
the total spin (from $S_{\rm T} \approx 0.0$ to $\approx 3/4$) as $t^{\prime \prime}$ decreases.
The two main peaks in the conductance discussed up to now were the 
$S_{\rm T}=1$ Kondo peak at 1/4-filling (relevant in the strong inter-dot tunneling regime) 
and the peak at 1/2-filling (relevant in the weak inter-dot tunneling regime).
It is interesting to discuss how these peaks evolve as 
$\Delta V$ increases. Fig. 4a shows results for 
$t^{\prime \prime}/t^{\prime} \gg 1$ ($t^{\prime \prime}=1.0$, $U=5.0$ and $t^{\prime}=0.2$).
The solid line displays the conductance and the doted line displays the total spin 
$S_{\rm T}$. Level separation $\Delta V$ increases from bottom 
to top (values for each graph are displayed in the left side). 
From $\Delta V=0.0$ up to $\Delta V \approx 0.6$ the width of the 
conductance peak slowly decreases, as also does the maximum value of $S_{\rm T}$. 
Above $\Delta V \approx 0.7$ (not shown) the narrowing of the peak accelerates 
(as does the decrease of $S_{\rm T}$), until the peak has all but vanished for $\Delta V = 1.0$. 
For higher values of $\Delta V$ (top graph, $\Delta V=10.0$), the 
conductance  shows the typical Coulomb blockade profile previously discussed \cite{carlos1}
for coupled single-level dots when $t^{\prime \prime}/t^{\prime} \gg 1$.

\begin{figure}[h]
\centering
\includegraphics[width=7.0cm]{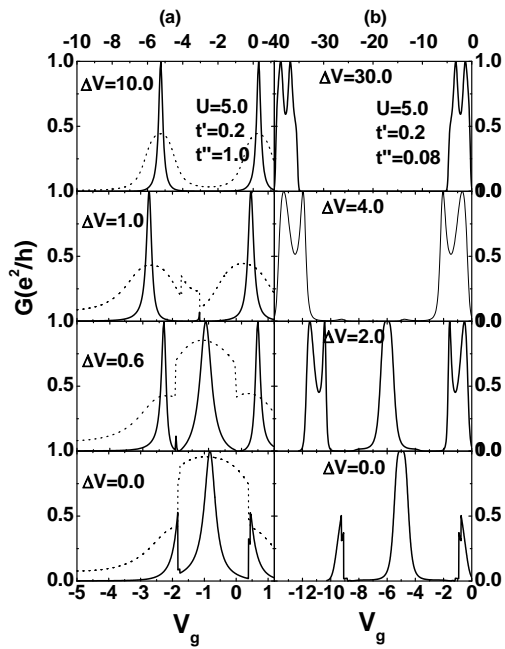}
\vspace{-0.5cm}
\caption{ Variation of the conductance with level separation $\Delta V$. 
The level separation is indicated for each curve. (a) $t^{\prime \prime}=1.0$:
The gate potential decreases down to the particle-hole symmetric value, to highlight how the 
$S_{\rm T}=1$ Kondo peak varies with $\Delta V$. Note that all graphs have the same 
horizontal axis scale, except for the upper one 
(scale indicated on top).
(b) $t^{\prime \prime}=0.08$:
The gate potential varies down to the lowest value (total charging of the dots), to highlight
the variation of the central peak (at the particle-hole symmetric point). Here also 
the upper graph has a different horizontal scale. A discussion of how the data interpolate between 
double- and single-level dots is given in the text.
}
\label{fig4}
\end{figure}
Figure 4b shows the corresponding results for $t^{\prime \prime}/t^{\prime} < 1$
($t^{\prime \prime}=0.08$, $U=5.0$ and $t^{\prime}=0.2$). Note that 
the central peak does not change appreciably from $\Delta V=0.0$ to $\Delta V=2.0$. In fact, 
changes start only above $\Delta V=3.0$, when  
the central peak splits in two ($\Delta V \gtrsim 3.2$, not shown). For $\Delta V > 3.4$ 
the two peaks start moving farther apart from each other and become very narrow. Finally, for 
$\Delta V=4.0$ the central peaks have disappeared, and the remaining structures are already 
similar to the single-level result $t^{\prime \prime}/t^{\prime} < 1$. The top graph 
($\Delta V =30.0$) is basically the result reported for single-level 
QDs at weak interdot tunneling \cite{carlos1} (if one discards the slight shoulders 
in the internal peaks).

Based on the results displayed on Fig. 4a, a qualitative phase diagram for 
the strong inter-dot tunneling regime can be sketched. In Fig. 5, the 
electron occupancy is in the 
horizontal axis (controlled by $V_g$) and $\Delta V$ is in the vertical 
axis. The left side (indicating 1/2-filling) is dominated by antiferromagnetism for 
all values of $\Delta V \lesssim U$. The singlet formed by the four levels is made of two 
spins $S \approx 1.0$. For $\Delta V > U$ one recovers the single-level picture.
The right side of the phase diagram, which describes the evolution 
of the central peak in Fig. 4a, is more interesting. For $\Delta V < t^{\prime \prime}$ 
one has the novel FM region, where an $S_{\rm T}=1$ Kondo effect is present. For $\Delta V > 
t^{\prime \prime}$ an AF region is found, with no Kondo effect.

\begin{figure}[h]
\centering
\includegraphics[width=5.0cm]{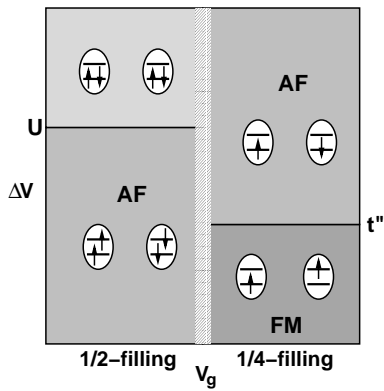}
\caption{ Qualitative phase diagram for the strong inter-dot tunneling regime ($t^{\prime \prime}/t^{\prime} \gg 1$). 
Horizontal axis indicates the level occupancy (controlled by  $V_g$) and vertical axis indicates 
the level separation (controlled by $\Delta V$). In the 1/2-filling region, 
displayed in the left side, one goes from an AF coupling between 
two $S=1$ spins (for $\Delta V < V_g$) to a situation where the two lower 
levels are completely occupied (for $\Delta V > V_g$). In both cases there is 
no Kondo effect. On the other hand, in the 1/4-filling region (right side), 
for $\Delta V < t^{\prime \prime}$ one has the novel $S_{\rm T}=1$ FM Kondo region, which gives 
way to an AF region (with no Kondo effect) once $\Delta V > t^{\prime \prime}$. 
The regime with 3 electrons in the two dots is very narrow as a function of $V_g$ and it is not shown.
}
\label{fig5}
\end{figure}

A finite size scaling analysis was done (results not shown) to verify how our 
numerical results converge 
with cluster size\cite{finite}. The authors found out very little change in the 
results with increasing cluster size, giving us confidence
that all the qualitative results here discussed are not caused by finite-size
effects. It is also important to stress that the calculations presented in Fig. 2a 
were reproduced for $U^{\prime}=J=0.0$ (with the values of all other parameters kept the same as 
before ($U=5.0$, $t^{\prime \prime}=1.0$ and $t^{\prime}=0.2$)) and 
the results obtained barely changed. This indicates that the FM correlation and
the $S_{\rm T}=1$ Kondo should be experimentally observable, since the only requirement is to have
two double-level QDs with strong inter-dot tunneling\cite{note3}.

Figure 6 qualitatively summarizes the main result presented in this Letter: (a) For 
double-level coupled quantum dots in the strong inter-dot tunneling regime at 1/4-filling, 
FM correlations will develop and conductance through a Kondo channel is allowed. 
(b) On the other hand, single-level coupled QDs will develop AF correlations in the 
strong inter-dot tunneling regime and conductance is suppressed.
The results discussed in this paper complete the analogy between QD states and 
magnetic phenomena in bulk materials. Previous investigations had shown that Kondo 
and AF states were possible in QDs. Now, at least theoretically, a regime with ferromagnetism 
has also been found, if more than one level per dot is active. Certainly, it would be important 
to confirm experimentally this prediction. Our calculations emphasizing multilevel dots present 
analogies with multi-orbital materials such as manganites, nickelates, cobaltites, and ruthenates. 
These compounds have a plethora of phases, all of which could find realizations in QDs systems 
as well.

\begin{figure}[h]
\centering
\includegraphics[width=5cm]{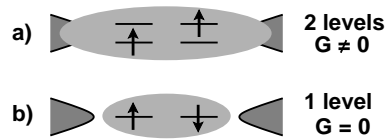}
\caption{ Schematic representation of the main result in this Letter}
\label{fig7}
\end{figure}

The authors acknowledge conversations with E. V. Anda and G. Chiappe. Support was provided 
by the NSF grants DMR-0122523 and 0303348.


\begin{thebibliography}{}
\bibitem{note1} $G_{lr}(E_{\rm F})$ is the Green function that
moves an electron from the left to the right lead, $E_{\rm F}$ is the
Fermi energy, and $\rho(E_{\rm F})$ is the density of states of
the leads:
$\rho(\omega) = \frac{-1}{\pi} \mbox{Im} \left\{
\frac{(\omega+i\eta) - \sqrt{(\omega+i\eta)^2 - 4t^2}}{2t^2}
\right\}$.
\bibitem{note2} Note that the total spin in the four levels in the two 
quantum dots (denoted as $S_T$ and obtained through $\langle \vec{S_{\rm T}}^2 
\rangle_{\rm cluster}$=$S_{\rm T}(S_{\rm T}+1)$, where 
$\vec{S_{\rm T}}=\sum_{i,\lambda}{\vec{S_{i,\lambda}}}$, where $i$ labels the dots 
and $\lambda$ labels the levels) is not a good quantum number 
(since the dots are not isolated). However, $S_{\rm T}$ gives a good indication 
on the nature of the Kondo effect ($S=1/2$ or $S=1$).
\end{thebibliography}
\end{document}